\documentstyle[aps,pre,epsfig,multicol,psfig]{revtex}
\begin{document}
\draft

\widetext

\title{Numerical simulation of vibrated granular gases under realistic
boundary conditions}

\author{Welles A. M. Morgado and Eduardo R. Mucciolo}
\address{Departamento de F\a'{\i}sica, Pontif\a'{\i}cia Universidade
Cat\'olica do Rio de Janeiro \\ CP 38071, 22452-970 Rio de Janeiro,
Brazil}

\date{\today}

\maketitle

\begin{abstract}
A variant of the Direct Simulation Monte Carlo method is used to study
the behavior of a granular gas, in two and three dimensions, under
varying density, restitution coefficient, and inelasticity regimes,
for realistic vibrating wall conditions. Our results agree quite well
with recent experimental data. The role of the energy injection
mechanism is discussed, as well as the behavior of state-functions,
such as pressure, under realistic boundary conditions. Upon a density
increase, we find signals of a clustering transition.
\end{abstract}


\section{Introduction}
\label{sec1}

The study of granular systems (GS) is one of the areas of modern
physics that has presented the fastest growth lately
\cite{jaeger96}. The interest in such systems ranges from purely
theoretical, as is the case of the self-organized critical sand-pile
model \cite{SOC}, to commercial, as in the applications oriented to
the pharmaceutical industry \cite{shinbrot1}. GS are composed of solid
matter in the form of irregularly shaped macroscopic grains. Their
dynamical and statistical properties can also be affected by the
presence of an interstitial medium, such as air or a liquid.

Three recent experiments dealt with two-dimensional granular gases
 \cite{herrmann01} subject to periodic shaking
 \cite{kudrolli1,menon1,kudrolli2}. It was observed that certain
 quadratic degrees of freedom, only indirectly coupled to the driving
 walls through the phase space randomization occurring during
 collisions, present non-Gaussian distributions in the steady
 state. Some of these distributions were nicely fitted by stretched
 exponentials \cite{menon1,losert99,olafsen98} with an exponent
 $\alpha = 3/2$, while others deviate altogether from the Gaussian
 behavior in the dense gas regime \cite{kudrolli2}. Non-Gaussian tails
 are indeed predicted by the kinetic theory of dissipative granular
 gases \cite{noje98}. Thus, these observations indicate that the
 dynamics of granular systems may depart strongly from that of
 ordinary gases which are well described by Gaussian distributions.


In this paper we study how inelasticity and boundary conditions
(either energy feeding or absorbing) affect the development of the
steady state of two and three dimensional granular gases. By employing
numerical simulations with realistic boundary conditions, we make
contact with the experimental settings
\cite{kudrolli1,menon1,kudrolli2}, studying their similarities with
respect to the energy transfer processes. Our numerical work is based
on a variant of the Direct Monte Carlo Simulation (DSMC) method
\cite{DSMC}. The implementation of this method is simple and yields
results in very good agreement with the experiments with essentially
no fitting parameters.

We find that velocity distributions and their moments depend strongly
on the energy feeding mechanisms, as well as on boundary
conditions. In our model, energy enters the system through vibrating
walls and exits through collisions, either among grains, or between
grains and fixed walls, which act as energy sinks. For the inelastic
granular gas, we find that the velocity distributions loose their
Gaussian character, as expected. (Recall that only for elastic systems
the Gaussian form may be associated with the fact that the kinetic
energy, a quadratic form of the velocities, is conserved.) In the
steady state, we also observe that the ratio between mean square
velocities along perpendicular directions is not unit. Nevertheless,
this ratio is found constant over several decades of vibrating wall
velocities. As we show below, this is a consequence of energy entering
the system from a privileged direction. Finally, for any inelasticity
coefficient value, as the density of the gas is increased, an
instability towards the formation of clusters appears. While our
simulations are not quantitatively reliable in the high-density limit,
they do provide a way of characterizing the tendency to
clusterization, intrinsic to dissipative granular systems.

Other recent DSMC simulation \cite{puglisi98,BMPV} have addressed the
problem of a granular gas under the influence of a gravitational
field. In contrast, we use a different form of DSMC simulation,
including some aspects of grain-wall and grain-grain collisions that
were not taken into account before, and focusing on the effect of
energy feeding and dissipation. In addition, We also study how the
lateral walls affect the internal energy balance. Moreover, our
implementation of multiple grain-collision has an alternative,
stochastic form.

The remaining of the paper is organized as follows. In Section
\ref{sec2}, we describe the granular system we will study. In Section
\ref{sec3}, we derive a convenient form for the one-particle density
distribution used in the model. In Section \ref{sec4}, we define the
length and time scales relevant to the problem. In Section \ref{sec5},
we calculate the pair collision probability. In Section \ref{sec6}, we
sketch the main steps for the computational implementation of the
model. In Section \ref{sec7}, we show in detail the boundary
conditions employed in our version of the DSMC simulation. In Section
\ref{sec8}, we present our numerical results. In Section \ref{sec9},
we discuss our results and conclude.


\section{Description of the system}
\label{sec2}

The main object of our study is the complex behavior of granular
gases, especially that induced by inelasticity and fluctuations. The
simplified models we are going to analyze are systems composed of
$N\gg1$ smooth, spherical grains of the same diameter $d$ and mass
$m$. We suppose the system to be contained in a two-dimensional
rectangular box of lengths $L_x$ and $L_y$ (later we will generalize
the model to three dimensions). Rolling, as well as static friction
between the grains and the two-dimensional plane are neglected.

For smooth grains, the interaction only happens upon collisions and in
the radial direction. This force is composed of a conservative,
Hertzian, elastic term and a radial friction one \cite{brilliantov86},
\begin{equation}
{\bf F}_{12} ({\bf r}_{12}, {\bf c}_{12}) = k\, (r_{12} -
d)^{\frac{3}{2}}\, \hat{\bf r}_{12} - F_{\rm inel} (r_{12}, {\bf
c}_{12} \cdot \hat{\bf r}_{12} )\, \hat{\bf r}_{12},
\end{equation}
where ${\bf r}_{12}$ is the vector connecting the centers of mass of
grains 1 and 2 and ${\bf c}_{12}$ is their relative velocity. The
friction term is responsible for the loss of mechanical energy that
occurs during a granular collision. In general, that loss is well
characterized by a velocity- and material-dependent coefficient of
restitution $\epsilon({\bf c}_{12} \cdot \hat{\bf r}_{12})$.

Due to the continuous collisional loss of energy, one needs to inject
kinetic energy into the system to prevent it from collapsing. In the
experiments \cite{kudrolli1,menon1,kudrolli2}, this is done by means
of rapidly oscillating walls, with frequency $\nu$ and amplitude $A$,
at opposite extremes of the box. Here we take the limits $\nu
\rightarrow \infty$ and $A \rightarrow 0$, keeping $A \nu$
constant. That corresponds to the regime exploited in the
experiments. In this manner, energy is given to the system
incoherently through collisions with the moving walls. Thus, the
system reaches a steady state dependent only on $\nu A$, and
$\epsilon$. In this case, the granular gas can be characterized by its
phase-space density distribution.


\section{Density distribution}
\label{sec3}


\subsection{Discretization of the distributions}

In general, we do not have the complete information about all
positions and velocities of the system. It is usually more appropriate
to assume that the one-particle distribution characterizes a
stochastic process which can be better described by $M\geq N$
``quanta" of density. These quanta will represent a smooth
distribution in phase space. The smooth one-particle density
distribution can be written in terms of the $M$ quanta as
\begin{eqnarray}
f({\bf r},{\bf c},t)(\Delta c)^2 a^2 & = & \alpha({\bf c}, \Delta c;
{\bf r}, a^2)\, f_0.
\end{eqnarray}
The normalization for the two-dimensional distribution is then given
by
\begin{eqnarray*}
 \int_{\cal A} d^2r \int_{-\infty}^{\infty} d^2c\ f({\bf r},{\bf c},t)
 & = & \sum_{\rm phase\ space} f({\bf r},{\bf c},t) (\Delta c)^2
 a^2,\\ \Rightarrow N & = & Mf_0,\\ \Rightarrow f_0 & = & \frac{N}{M},
\end{eqnarray*}
where ${\cal A}$ denotes the area confining the grains and we used the
fact that $\alpha({\bf c},\infty; {\bf r},{\cal A}) = M$. Thus, a
convenient form for the two-dimensional $f({\bf r},{\bf c},t)$ is
\begin{equation}
f({\bf r},{\bf c},t) = \frac{N}{M a^2 (\Delta c)^2}\, \alpha (\Delta
c,a).
\label{f.2D}
\end{equation}
Similarly, we may write for the three-dimensional gas
\begin{equation}
f({\bf r},{\bf c},t) = \frac{N}{M a^3 (\Delta c)^3}\, \alpha( \Delta
c,a).
\label{f.3D}
\end{equation}
In the following, we proceed to treat the two and three dimensional
granular gas in a discretized form.


\subsection{Two-dimensional gas}

The total density distribution variation, due to collisions between
grains with velocities ${\bf c}^\prime$ and ${\bf c}$, after a time
interval $\delta t$ is given by \cite{chapman}
\begin{equation}
\delta f({\bf c},{\bf c}^\prime) = f({\bf c})\, f({\bf c}^\prime)
|{\bf c}^\prime - {\bf c}| (\Delta c)^2 d\, \delta t.
\label{varia.f2}
\end{equation}
We may insert Eq. (\ref{f.2D}) into Eq. (\ref{varia.f2}) and obtain
\begin{equation}
\delta f({\bf c},{\bf c}^\prime) =  \alpha(\Delta c,a)\,
\alpha(\Delta c^\prime,a) \frac{N^2 d\, |{\bf c}^\prime - {\bf c}|\,
\delta t} {M^2 a^4 (\Delta c)^2}.
\label{1amaneira.2D}
\end{equation}
It is important to remark that the molecular chaos hypothesis for the
collisional probability holds well for an inelastic granular system in
the the gaseous phase \cite{soto01}.


\subsection{Three-dimensional gas}

Similarly, for the three-dimensional case we have
\begin{equation}
\delta f({\bf c},{\bf c}^\prime) = \pi\, f({\bf c}) f({\bf c}^\prime)
|{\bf c}^\prime-{\bf c}| (\Delta c)^3\, d^2\, \delta t.
\label{varia.f3}
\end{equation}
Inserting Eq. (\ref{f.3D}) into Eq. (\ref{varia.f3}) we obtain
\begin{equation}
\delta f({\bf c},{\bf c}^\prime) =  \alpha(\Delta c,a)\,
\alpha(\Delta c^\prime,a) \frac{N^2 \pi\, d^2\, |{\bf c}^\prime - {\bf
c}|\, \delta t} {M^2 a^6(\Delta c)^3}.
\label{1amaneira.3D}
\end{equation}
%


\section{Natural scales and dimensionless units}
\label{sec4}

There are two important length scales in this problem: the grain
diameter $d$ and the box length $L$ (for simplicity, let us assume a
square box hereafter). We choose to express all lengths in units of
$d$. We also choose a suitably short time interval $\delta t$ as our
time scale. This $\delta t$ will correspond to the computational time
step in real, physical, terms. All lengths are scaled by $d$ and all
times by $\delta t$ in such a way that (the asterisks denote rescaled
quantities)
\begin{equation}
 f({\bf r},{\bf c},t) = f^\ast({\bf r}^\ast,{\bf
c}^\ast,t^\ast)\, \left( \frac{\delta t} {d^2} \right)^D,
\end{equation}
where $D$ is the spatial dimension. In terms of dimensionless
variables, Eqs. (\ref{f.2D}) and (\ref{1amaneira.2D}) become
\begin{equation}
f^\ast ({\bf r}^\ast,{\bf c}^\ast,t^\ast) = \frac{N} {Ma^{\ast\,
2}(\Delta c^\ast)^2} \, \alpha(\Delta c^\ast,a^\ast),
\label{flinha.2D}
\end{equation}
and
\begin{equation}
\delta^2 f^\ast({\bf c}^\ast,{\bf c}^{\prime\, \ast}) = \alpha
(\Delta c^\ast,a^\ast)\, \alpha(\Delta c^{\prime\, \ast},a^\ast)
\frac{N^2|{\bf c}^{\prime \ast} - {\bf c}^\ast|} {M^2 a^{\ast\,
4}(\Delta c^\ast )^2},
\label{1aman.2D}
\end{equation}
respectively. Similarly, the three-dimensional Eqs. (\ref{f.3D}) and
(\ref{1amaneira.3D}) become
\begin{equation}
f^\ast({\bf r}^\ast,{\bf c}^\ast,t^\ast) = \frac{N}{Ma^{\ast\,3}
(\Delta c^\ast)^3}\, \alpha(\Delta c^\ast,a^\ast),
\label{flinha.3D}
\end{equation}
and
\begin{equation}
\delta^3 f^\ast({\bf c}^\ast,{\bf c}^{\prime\, \ast}) =  \alpha
(\Delta c^\ast,a^\ast)\, \alpha (\Delta c^{\prime\, \ast},a^\ast)
\frac{N^2 \pi |{\bf c}^{\prime \ast} - {\bf c}^\ast|} {M^2 a^{\ast\,
6} (\Delta c^\ast )^3},
\label{1aman.3D}
\end{equation}
%


\section{Calculation of collision probabilities}
\label{sec5}

Another way of obtaining Eqs. (\ref{1aman.2D}) and (\ref{1aman.3D}) is
to calculate the variation of $f^\ast({\bf r}, {\bf c}, t)$ due to the
collisional dynamics inside phase space regions of volume $a^2(\Delta
c)^2$ and $a^3(\Delta c)^3$, respectively. That variation happens in
small jumps due to the collisions that actually happen between all
pairs of quanta. The total change, in two dimensions, is given by
\begin{equation}
\delta^2 f({\bf c},{\bf c}^\prime) = -\alpha(\Delta c,a)\,
\alpha(\Delta c^\prime,a) \frac{N}{M a^2 (\Delta c)^2}\, p^{\rm
2D}_{{\rm col},\delta t}.
\label{2aman.2D}
\end{equation}
A comparison between Eqs. (\ref{2aman.2D}) and (\ref{1amaneira.2D})
yields the pair collision probability
\begin{eqnarray}
p^{\rm 2D}_{{\rm col},\delta t} & = & \frac{N |{\bf
c}^{\prime\, \ast} - {\bf c}^\ast|}{M a^{\ast\, 2}}
\label{pcol2}
\end{eqnarray}
in dimensionless units. Similarly, we obtain, for the
three-dimensional case,
\begin{equation}
p^{\rm 3D}_{{\rm col},\delta t} = \frac{N\pi |{\bf c}^{\prime\, \ast}
- {\bf c}^\ast|} {M a^{\ast\, 3}}.
\label{pcol3}
\end{equation}
Notice that the pair collision probability of density quanta goes as
$M^{-1}$, decreasing as $M \rightarrow \infty$. However, the collision
rate for the actual system remains constant, since the decrease in
$p_{\rm col}$ is compensated by the increase in the number of
colliding quanta, which goes as $M^2$, and a decrease of the
normalization factor (which varies as $M^{-1}$).

In the next Section, we introduce the Direct Simulation Monte Carlo
(DSMC) model \cite{DSMC}, which, in the appropriate limit of $N$ fixed
and $M \rightarrow \infty$, reproduces the usual kinetic theory model.


\section{The DSMC and time evolution}
\label{sec6}

In our model, space is discretized, but velocities are not. For
two-dimensional systems, we divide the position space into boxes of
volume ($a^3$ in three dimensions), and, inside each box, the
collisional dynamics for the density quanta happens as in the usual
low density limit. A streaming density flux is established among
neighboring cells. We allow for the presence of external fields, such
as gravity, by the inclusion of an external impulse given to each
grain at each time step. In the following, we describe the DSMC
\cite{DSMC} and its evolution.

Our version of the DSMC consists of $M$ density quanta (DQ), each
carrying the contribution $\frac{N} {M a^2(\Delta c)^2}$ [or,
$\frac{N} {M a^3(\Delta c)^3}$ in the three-dimensional case], located
in cells of area $a^2$ (volume $a^3$ in three dimensions). They
reproduce exactly the smooth density distribution $f({\bf r}, {\bf c},
t)$ in the limit $M \rightarrow \infty$ and $a \rightarrow 0$ for a
fixed $N$.

The time evolution is obtained by following the steps shown below:
\begin{itemize}
\item {\em Step 1:} Running over all cells, one constructs an ordered
list of all DQ inside each cell. For every DQ pair in a given cell,
one draws a random number in order to check whether a collision
happens or not, with probabilities calculated by Eqs. (\ref{pcol2}) or
(\ref{pcol3}). If the collision happen, both DQ are taken out of the
list, their new velocities calculated by drawing an impact parameter
$b$ (and an additional angle $\varphi$, in the three-dimensional case)
corresponding to the collision, and proceeding to the next pair in the
list. Any grain collide no more than once per iteration.
\item {\em Step 2:} The streaming flux of quanta is calculated for
each DQ in an cell. For instance, let us take a quantum located at
position $(k_x,k_y)$, with $k_x,k_y$ integers, and velocity
$(c_x,c_y)$. Thus, the distribution function can be written as
\[
f^{\ast}\equiv f^{\ast}(k_x,k_y,c_x^{\ast},c_y^{\ast},t^{\ast}).
\]
Its new position in the $x$-direction is obtained by drawing a random
number $\eta$ in the interval $[0;1]$: it will be $k_x + [c_x \delta
t/a]$ if $\eta < [c_x\delta t/a]$, where $[z]$ means the integer part
of $z$; it will be $k_x + [c_x\delta t/a] + 1$ if $\eta \geq
[c_x\delta t/a]$. The analog operation is carried out for the flow
along $y$. The presence of an external field {\bf g}, such as gravity,
can be taken into account by adding the impulse
\[
{\bf c}\rightarrow {\bf c}+{\bf g}\, \delta t,
\] 
to the particles at this step \cite{grav}.
\item {\em Step 3:} The walls' boundary conditions are applied (see
below).
\end{itemize}
After completing {\em step 3} one returns to {\em step 1}. In the
following, we explain in detail the computer implementation of the
DSMC model.


\section{Algorithm for the DSMC}
\label{sec7}


\subsection{Initial procedures}

We start by setting the initial distribution uniform in space and
Gaussian on the velocities: $\ln f_1(k_x, k_y, c_x, c_y) \propto -{\bf
c}^2$.


\subsection{Iteration at the borders -- boundary conditions and 
auxiliary sites}

To incorporate the effect of shaking into the boundary conditions, we
only need to consider grains whose position and velocity direction
indicate the possibility of an eminent collision with one of the
moving walls. For instance, let $- c_y < 0$ be the grain velocity
along $y$ for a grain located in a cell of $y=1$ coordinate at a given
instant $t$. If not disturbed, such grain may collide with the moving
wall located at $y=0$ (the bottom wall) within an interval $\delta
t$. If the streaming drawing decides in favor of the collision, the
following procedure is adopted. We approximate the wall motion by a
sawtooth oscillation with velocities $\pm W_b$. The amplitude $A$ and
frequency $\nu$ are related by $W_b = A \nu$, where we take the limit
$A\rightarrow 0$ and $\nu\rightarrow \infty$. Also, we allow for the
grain-wall collision to be inelastic by introducing the inelastic
coefficient $\epsilon_w$.

There are three possible ways for a grain to collide with a moving
wall. First, it can collide frontally, and only once, when the wall
has a positive velocity $+W_b$. In this case, the grain final velocity
is given by
\begin{equation}
c_+ = W_b(1 + \epsilon_w) + c_y\epsilon_w.
\end{equation}
Second, it can also collide only once when the wall has a negative
velocity $-W_b$. In this case, we have
\begin{equation}
c_- = -W_b(1 + \epsilon_w) + c_y \epsilon_w.
\end{equation}
Finally, the grain can collide twice: if the wall velocity is $-W_b$
at the moment of the collision and, afterwards, the grain velocity is
sufficiently reduced, the wall can hit the grain again. The second
collision is frontal. In this case, the grain final velocity will be
\begin{equation}
c_{-+} = W_b(1 + \epsilon_w)^2 - c_y \epsilon_w^2.
\end{equation}

We now proceed to calculate the probabilities $p_-,p_+,p_{-+}$ for
each kind of collision. Their expressions depend on which range of
velocities $c_y$ falls. Notice that $c_y$ is the absolute value of the
granular velocity near the wall. Let us define the following useful
velocities:
\begin{eqnarray*}
c_{1b} & = & W_b, \\
c_{2b} & = & W_b (1+1/\epsilon_w),\\
c_{3b} & = & W_b (1+2/\epsilon_w).
\end{eqnarray*}
The probabilities follow from a straightforward accounting of the
possible wall positions when a grain enters the region of size $2A$
centered at the average position of the oscillating wall. Once we
determine the type of collision, the next step is to calculate the
corresponding wall reflection flow. For each $-c_y$, the fraction of
grains colliding with the wall is given by $\theta = c_y \delta
t/a$. For cells neighboring the wall, the grain-wall probability
collision is
\[
\theta \times {\rm  corresponding\; probability} \times p_{-,+,-+}.
\]
Thus, for the DSMC model, each grain near the wall is tested by the
drawing of two random numbers. The first determines whether the grain
collides with the wall with probability $\theta$. The second
determines which kind of collision it will be ($c_{-,+,-+}$). Due to
the infinite frequency of wall oscillations, the grain-wall collisions
are completely uncorrelated. The grain-wall collision probabilities
are given in the following.

\begin{enumerate}

\item Single head-on collision ($0 < c_y < c_{1b}^L$): $p_+ = 1$.

\item Single head-on collision, or a head-tail collision, followed by
a head-on collision ($c_{1b} < c_y < c_{2b}$): $p_+ = \frac{1}{2}
\left( 1 + \frac{W_b}{c_y} \right)$ and $p_{-+} = \frac{1}{2} \left( 1
- \frac{W_b}{c_y} \right)$.

\item Single head-on collision, or a head-tail collision, followed by
a head-on collision, or a single head-tail collision ($c_{2b} < c_y
\le c_{3b}$): $p_+ = \frac{1}{2} \left( 1 + \frac{W_b}{c_y} \right)$,
$p_- = 1 - \frac{W_b}{c_y} \left( 1 + \frac{1}{\epsilon_w} \right)$,
and $p_{-+} = \frac{1}{2} \left[\frac{W_b}{c_y} \left( 1 +
\frac{2}{\epsilon_w} \right) - 1 \right]$.

\item Single head-on collision, or single head-tail collision ($c_y >
c_{3b}$): $p_+ = \frac{1}{2} \left( 1 + \frac{W_b}{c_y} \right)$ and
$p_- = \frac{1}{2} \left( 1 - \frac{W_b}{c_y} \right)$.

\end{enumerate}

We repeat the procedure above for the top wall, changing signs, site
coordinate, and wall velocity accordingly. The left and right walls
are assumed fixed.


\subsection{Velocity scale}

The computational time-scale is set by choosing
\begin{equation}
W_{b,t}^\ast = W_{b,t} \frac{\delta t}{d}.
\end{equation}  
for given experimental values of $W_{b,t}$ and $d$. For the case of
recent experiments on two dimensional granular gas
\cite{kudrolli1,menon1,kudrolli2}, the wall velocities were of the
order of 1 m/s, and $d \approx \times10^{-3}$m. For a computational
$W_{b,t}^\ast$ of the order of 1, we obtain $\delta t \approx
10^{-3}$s.


\section{Numerical results}
\label{sec8}

The main motivation for using the DSMC model is to explore, in detail,
the behavior of dilute granular gases, such as those studied in
Refs. \cite{kudrolli1,menon1}. Below, we list the main results
obtained from the DSMC. All quantities presented are rescaled
according to procedure of Sec. \ref{sec4} (asterisks are dropped).

When averaging over time, we used data points obtained sampling the
time sequence at every 10 to 20 steps, according to the system size,
in order to assure statistical independence. In what follows, we make
$M = N$.


\subsection{Vibrated two-dimensional gas}

\begin{figure}
\epsfxsize=6cm
\hspace{.5cm}
\vspace{.5cm} 
\epsfbox{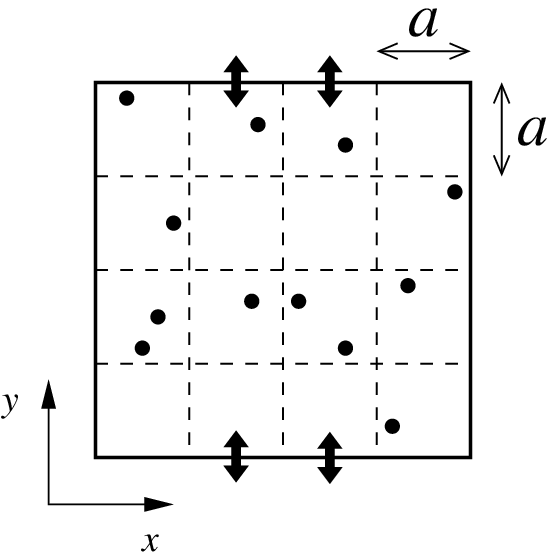} 
\caption{Schematic view of a two-dimensional, vertical granular
system. Spatial discretization is implemented by dividing the square
box containing the spherical grains into equal cells of side $a$. The
double vertical arrows indicate the position of the moving walls.}
\vspace{.5cm}
\label{fig1}
\end{figure}

\begin{figure}
\epsfxsize=8cm
\epsfbox{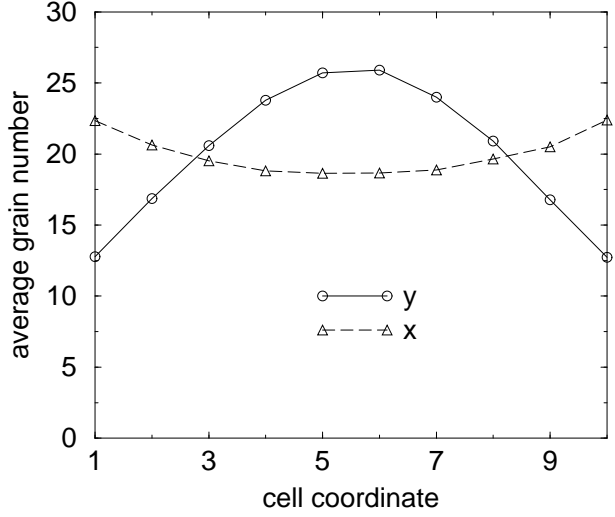} 
\caption{Grain density distributions for $10 \times 10$ cells, $a =
5$, $N = 200$, $\epsilon = 0.5$, $\epsilon_w = 0.8$, $W_b = W_t = 1$,
and $g = 0$.}
\vspace{.5cm}
\label{fig2}
\end{figure}

The scheme is presented in Fig. \ref{fig1}. It mimics the experimental
setting of Rouyer and Menon \cite{menon1}. Their results for the
steady-state grain densities profile can be reproduced quantitatively
by our numerical simulations when we chose numerical values for grain
number, aerial volume fraction, and inelastic coefficients equivalent
to the experimental ones. For instance, Fig. \ref{fig2} shows the
vertical (longitudinal) and horizontal (transverse) average grain
number per cell density profiles for the two-dimensional square
box. It is important to remark that the wall rms acceleration in
Ref. \cite{menon1} was sufficiently strong to render the gravitational
field ineffective. We thus have set $g=0$ in these simulations. Notice
that the grain density along $y$ is lower near the moving walls,
reaching a maximum in the middle of the box. The opposite behavior
happens horizontally, although the density variation is less
pronounced. The velocity distributions $P(c_x)$ and $P(c_y)$ were
evaluated using only data from cells of maximum density along $y$,
namely, $k_y=5,6$ for $10 \times 10$ cells.

\begin{figure}
\epsfxsize=8cm
\epsfbox{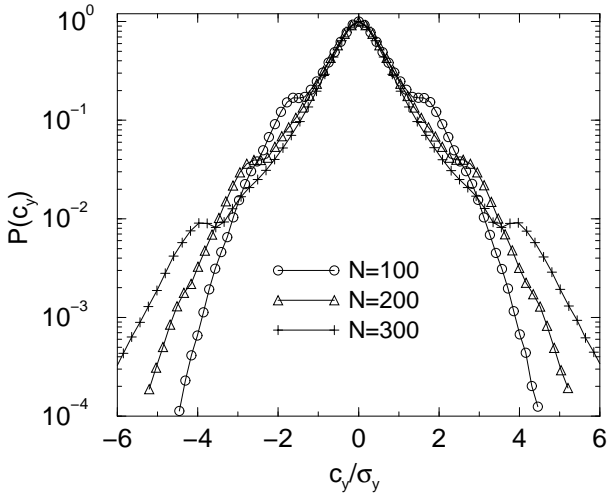}
\caption{The vertical (longitudinal) velocity distributions for the
same parameters as in Fig. \ref{fig2}. The solid lines are only guides
for the eyes.}
\vspace{.5cm}
\label{fig3}
\end{figure}

In Fig. \ref{fig3}, we show the velocity distribution along the
direction (vertical) parallel to the moving walls, $P(c_y)$, for a
total grain number ranging from $N=100$ to $300$. To facilitate the
comparison, the velocities are rescaled by their mean square value,
$\sigma_y = \langle c_y^2 \rangle^{1/2}$. These curves agree
qualitatively with the experimental results, showing the symmetric
shoulders characteristic of the energy injection mechanism. In fact,
for a given grain number, each distribution is formed by the
superposition of three identical curves: one centered at $c_y=0$ and
two others centered at $c_y = \pm c_y^w$, where $c_y^w$ increases with
the wall velocity and decreases with $\epsilon_w$. Notice that the
shoulders do not coincide because the rescaling factor $\sigma_y$
varies with the grain number. Another interesting feature is the
nearly exponential tails. Similar results were obtained by Baldassari
{\it et al.} in another recent DSMC simulation for the horizontal
distributions \cite{BMPV}.

\begin{figure}
\epsfxsize=8cm
\epsfbox{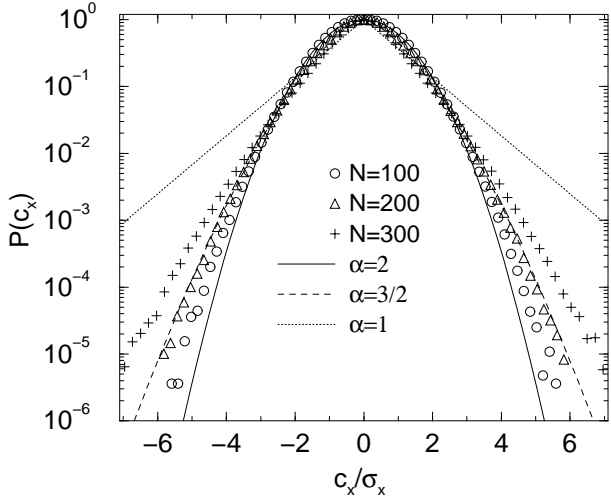}
\caption{The horizontal (transversal) velocity distributions for the
same parameters as in Fig. \ref{fig2}. The exponent $\alpha$ indicates
exponential (1), stretched exponential (3/2), and Gaussian (2)
curves.}
\vspace{.5cm}
\label{fig4}
\end{figure}

\begin{figure}
\epsfxsize=8cm
\epsfbox{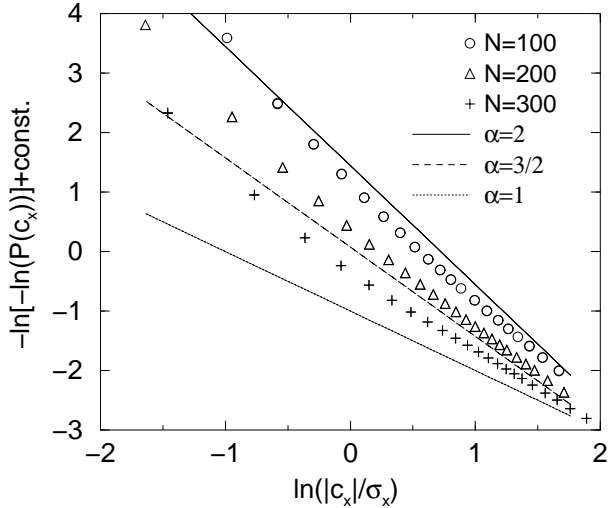}
\caption{The tails of velocity distributions shown in
Fig. \ref{fig4}.}
\vspace{.5cm}
\label{fig5}
\end{figure}

Figure \ref{fig4} shows the horizontal velocity distribution $P(c_x)$
for the same grain numbers of Fig. \ref{fig3} and a similar rescaling
($\sigma_x = \langle c_x^2 \rangle^{1/2}$). We observe that, as the
number of grains (and consequently the average number of collisions
per grain per unit of time) is increased, the form of the distribution
changes smoothly from nearly Gaussian to, approximately, an stretched
exponential of the form $A\, e^{-\beta\, |c_x/\sigma_x|^{\alpha}}$,
with a density-dependent exponent $\alpha$. For $N=200$, the exponent
$\alpha \approx 3/2$ fits approximately the tails of the distribution
(the normalization condition requires that $\beta \approx 0.797$ when
$A=1$ in this case). For $N=300$, $\alpha$ is certainly smaller than
$3/2$. This is illustrated in Fig. \ref{fig5}, where the tails of the
horizontal velocity distributions are presented in a log-log scale.

\begin{figure}
\epsfxsize=8cm
\hspace{.5cm}
\epsfbox{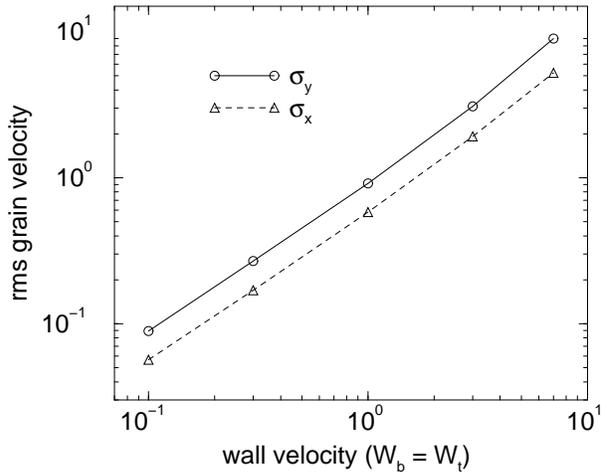} 
\caption{Grain root mean square velocity dependence on wall
velocities. The simulation parameters are the same in
Fig. \ref{fig2}. Units follow the rescaling introduced in
Sec. \ref{sec4}. Notice that $\sigma_x \neq \sigma_y$, but their ratio
is constant.}
\vspace{.5cm}
\label{fig6}
\end{figure}

We repeated the numerical simulations for several values of $W_b$,
$W_t$, and $\epsilon_W$. We verified that the only important aspect
about the velocity of the vibrating walls is the injection of energy
into the system that they are responsible for. The wall velocities and
wall-grain restitution coefficient set the steady-state average
velocity in the gas, as shown in Fig. \ref{fig6}, but do not alter the
shape of the (rescaled) velocity distributions. 

\subsection{Dependence on inelasticity}

\begin{figure}
\epsfxsize=8cm
\hspace{.5cm}
\epsfbox{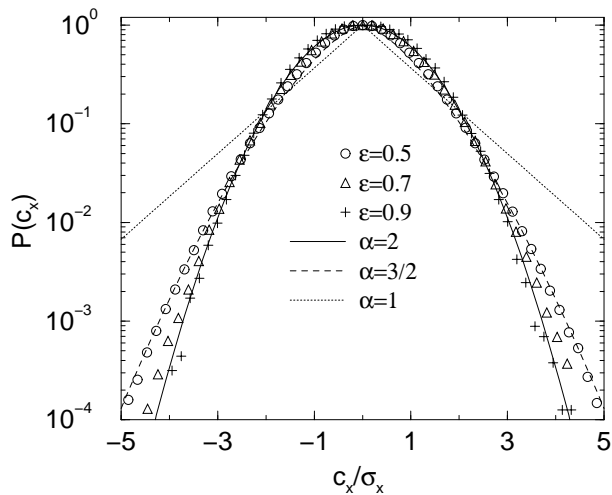} 
\caption{The dependence on $\epsilon$ of the horizontal velocity
distribution. The parameters are the same as in Fig. \ref{fig2}.}
\vspace{.5cm}
\label{fig7}
\end{figure}

While changes in the grain-wall restitution coefficient only rescale
the steady state average square velocity components, we found that the
grain-grain inelasticity does set the behavior of the velocities
distribution tails for sufficiently large and dense systems. That
effect can be seen in Fig. \ref{fig7}, where we present the
distribution of horizontal velocities for different values of
$\epsilon$. Notice that the distribution starts as Gaussian when the
inelasticity is weak, but evolves towards a stretched exponential as
$\epsilon$ becomes substantially smaller than unit.

\begin{figure}
\epsfxsize=8cm
\epsfbox{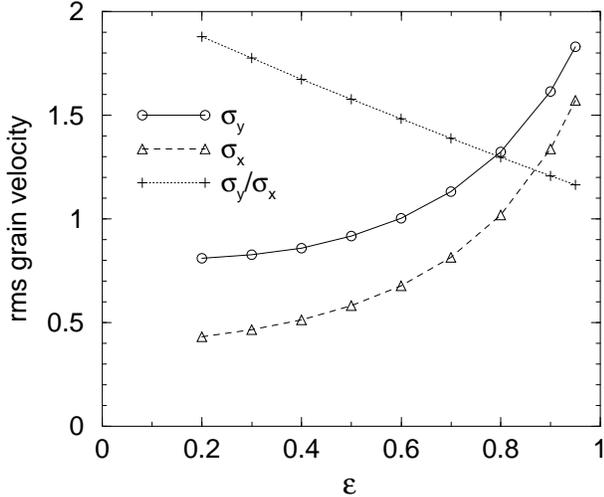} 
\caption{Grain root mean square velocity dependence on $\epsilon$
(inelasticity coefficient). The simulation parameters are same as in
Fig. \ref{fig2}, with $N=200$. Also shown is the ratio between the rms
velocity components. Notice that one does not expect
$\sigma_x/\sigma_y = 1$ in the elastic limit since energy enters from
the wall along $x$.}
\vspace{.5cm}
\label{fig8}
\end{figure}

\begin{figure}
\epsfxsize=8cm
\hspace{.5cm} 
\epsfbox{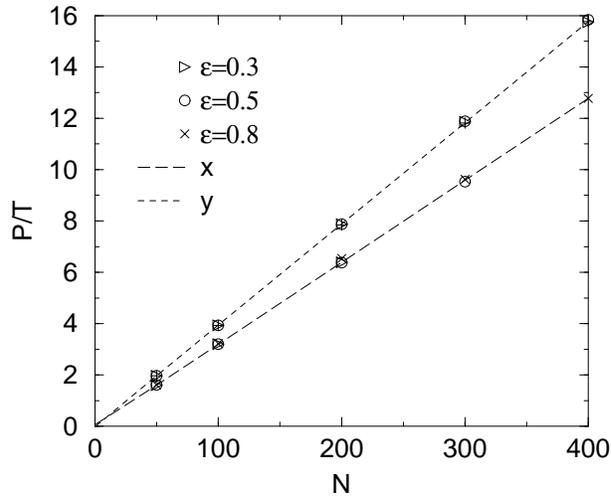} 
\caption{The ratio between wall pressure $P$ and temperature $T$ along
each direction as a function of grain number (fixed volume). The data
for different restitution coefficients are shown. Other simulation
parameters follow those of Fig. \ref{fig2}. All units follow the
rescaling introduced in Sec. \ref{sec4}.}
\vspace{.5cm}
\label{fig9}
\end{figure}

The influence of the grain-grain restitution coefficient in the
standard deviations of horizontal and vertical velocities are shown in
Fig. \ref{fig8}. Notice that the ratio $\sigma_y/\sigma_x$ approaches
1 as the system becomes nearly elastic. In the opposite limit, the
asymmetry between longitudinal and transverse average square velocities
increases as the system becomes more inelastic. The equipartition of
average kinetic energy between horizontal and vertical degrees of
freedom is therefore broken. In this case, it is meaningful to define
two granular temperatures for the system, namely,
\begin{equation}
T_x = \sigma_x^2 \qquad \mbox{and} \qquad T_y = \sigma_y^2.
\end{equation}
Pressure is defined on the wall as the average momentum transmitted to
it by the colliding grains per unit area and per unit time. Note that
pressure depends on the inelastic properties of the grains and the
wall. We thus also define $P_x$ and $P_y$ as the average momentum
transmitted for the horizontal and vertical walls, respectively. The
$PTN$ diagram for a fixed system area is shown in
Fig. \ref{fig9}. Notice that their interdependence is still linear
along a given direction. However, dynamical effects due to energy
injection mechanism appear related to the higher pressure on the
direction perpendicular to the moving walls.

We have also investigated a possible dependence of the $P(c_y)$ and
$P(c_x)$ on the functional form of $\epsilon (c_{12})$. The numerical
results (not shown) indicate that the influence of the detailed
dependence of $\epsilon$ on $c_{12}$ is weak when $\epsilon$ is not
too small.

\begin{figure}
\epsfxsize=8cm
\hspace{.5cm} 
\epsfbox{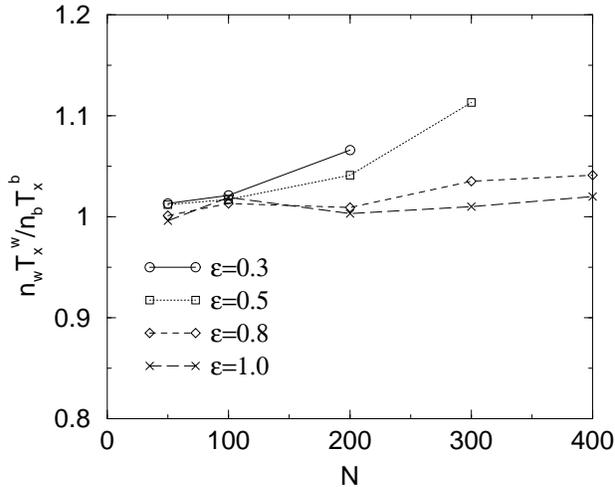} 
\caption{The ratio between $T_x \times n$ (the product of local
temperature and density) near the wall and in the bulk. While the
ratio remains very close to unit for quasi elastic systems, it begins
deviating strongly from that rule when the system is close to an
instability.}
\vspace{.5cm}
\label{fig10}
\end{figure}

The equation of state can be studied locally inside the gas volume
\cite{luding01}. In the steady-state regime, we found that the
granular temperature and the density vary inside the box, but their
product, proportional to the pressure, is constant along a fixed
direction. This can be seen in Fig. \ref{fig10}, where we have plotted
the ratio between the local quantity $T_x \times n$ near the wall and
in the bulk. However, we notice that for large inelasticities the
constancy of the product within the gas is no longer valid. In that
case, while pressure should still be homogenous along a given
direaction, the value obtained for the granular temperature is likely
to be incorrect. This occurs because we have ignored the multiple
grain-grain collisions at higher densities that should lower
considerably the value of the average kinetic energy. Therefore,
corrections associated with an increase in density are needed.

Furthermore, in the present model, we ignore the grain-wall tangential
friction. The presence of this type of friction in real systems leads
to cluster breakdown at the walls, thus reducing density instabilities
as those observed in our model. It also feeds energy into the degrees
of freedom parallel to the vibrating walls during grain-vibrating
walls collisions, helping to reduce the difference between $T_x$ and
$T_y$.

%
\subsection{Clustering transition}

The model fails when densities become too high. In that regime, the
particles tend to form clusters at low temperature and the collisional
probabilities used in the DSCM simulation becomes incorrect. In fact,
our method becomes unstable for high densities or strong
inelasticity. Such instabilities can be identified in the present
simulation by a sudden increase of granular density (sometimes of
several orders of magnitude) in certain regions of the box, especially
along the still walls. Thus, while the method does not allow any
quantitative description of the clusterized phase, it does provide a
way of identifying, within the molecular chaos hypothesis, a lower
bound for the onset of grain clusters \cite{pasini01,grossman97}.

\begin{figure}
\epsfxsize=8cm
\hspace{.5cm} 
\epsfbox{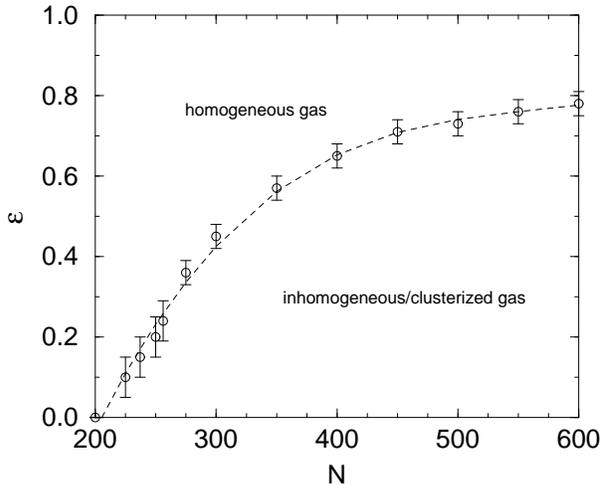} 
\caption{The value of $\epsilon$ below which an instability occurs for
a given $N$. Other parameters follow those of Fig. \ref{fig2}. The
dashed line is polynomial (cubic) curve fitted to the data points. For
$N<200$, no instability was found.}
\vspace{.5cm}
\label{fig11}
\end{figure}

The estimate an upper bound for the stability of the realistic
granular gas goes as follows. If an excessive density increase occurs
near a still wall (even at values so large that the excluded volume of
the particles would be larger than the region containing them), the
center-of-mass of the system will be dislocated towards that
wall. This provides us with a method to investigate the clustering
threshold: If the center-of-mass displacement is larger than a typical
fluctuative displacement, we check for excess density peaks near the
walls. If these are present, the model is assumed to fail in that
region. We thus plot a density-inelasticity phase diagram by marking
the upper boundary of stability obtained numerically (see
Fig. \ref{fig11}). We observe that for any value of the inelasticity,
there is a minimum value of the density for which a cluster appears
\cite{cluster}.

\subsection{The three-dimensional gas}

The generalization of the two-dimensional simulations to three
dimensions can be straightforwardly done in the DSMC context. We
investigated mainly two regimes (for not too small $\epsilon$),
namely, low and moderate densities. To facilitate the data
interpretation, we adopted hard wall boundary conditions and set
gravity equal to zero. The moving (equal velocity) walls operated in
the $x$ direction only.

\begin{figure}
\epsfxsize=8cm
\hspace{.5cm} 
\epsfbox{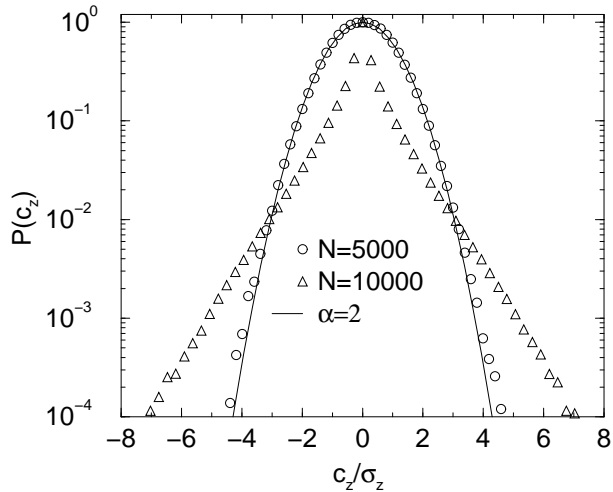} 
\caption{The distribution of velocity components perpendicular to the
moving walls ($z$ direction). System parameters are $10 \times 10
\times 10$ cells, $a = 5$, $W_b = W_t = 1$, $\epsilon = 0.5$,
$\epsilon_w = 0.8$, and $g = 0$. The distribution along $x$ is
essentially identical (not shown)}
\vspace{.5cm}
\label{fig12}
\end{figure}

For a small number of grains, $N = 5000$ in a cubic box of dimensions
$(5 \times d)^3$, the transverse velocities followed closely a
Gaussian distribution, as in the case of an elastic gas. This can be
seen in Fig. \ref{fig12}. The density distribution along the direction
parallel to the moving walls ($y$, not shown) is similar to the
two-dimensional case, having a maximum at zero velocity and two
lateral ``shoulders" whose positions are related to the vibrating wall
velocities, signaling that they are caused by the energy injection
mechanism.

The moderately dense gas case, with $N = 10,000$ in a cubic box of the
same volume, presents a much more interesting behavior. The velocity
distributions strongly depart from Gaussian, or even an stretched
exponential behavior, as seen in Fig. \ref{fig4}.

There seems to be no essential difference between the two- and the
three-dimensional granular gases. Our method allows us to investigate
the moderately dense granular gas limit, using pair collision
probabilities that are close to actual experimental values. We expect
to obtain reliable results as long as the probability of clustering is
small. We stress that velocities distributions seem to change
continuously from Gaussian to power-law behavior, as the effective
dissipation per grain per unit time is increased.


\section{Conclusions}
\label{sec9}

In this paper, we studied a two-dimensional granular gas via a
completely stochastic DSMC model. We used realistic boundary
conditions for energy feeding and energy absorbing walls. We analyzed
the behavior of the system for various regimes of granular densities
and inelasticity. We observed that the shape of the parallel
velocities distribution is highly dependent on the energy feeding
mechanism. The appearance of distinct granular temperatures for the
parallel and the transversal directions is one consequence of it. 

Our model reproduces experimental data for granular velocities
distributions on a satisfactory way, but fails whenever the density or
inelasticity take very large values. It also predicts a different set
of granular temperatures for the parallel and orthogonal directions to
the shaking walls. This is a manifestation of the fact that, for
granular inelastic systems on a steady-state, the equipartion theorem
of equilibrium statistical mechanics does not apply. A recent
experimental work has observed the same phenomenon for a mix of two
different granular systems \cite{menon_recent}.

We also studied some of the limits of such model by looking for the
breakdown of the gas phase due to the formation of clusters. We were
able to estimate an upper bound for the densities necessary for the
formation of clusters for a given inelasticity value. The type of
inelastic collapse associated to the formation of clusters reminds us
of experiments with vibrated granular systems where the bulk of the
system gets clusterized on one side of the volume, while the other
side remains in the fluidized phase \cite{shinbrot1,grossman97}.

The equation of state of a granular gas and the behavior of the $PTN$
diagram for moderate densities and inelasticities were
investigated. In principle, the onset of inelastic collapse could be
as well estimated by the behavior of the local density-temperature
product.

In summary, in this article, we present a computationally easy tool
for the study of the complex behavior of granular systems and study
some peculiar properties (e.g. equipartition breakdown, non-gaussian
behavior, etc.) and some of its limitations (e.g. clustering
formation).


\section{Acknowledgements}

We would like to thank CNPq, FAPERJ, and PRONEX for partial financial
support.



\end{document}